\documentclass{amsart}
\usepackage{amssymb,latexsym}
\usepackage{graphicx}

\begin{document}

\title{Integral equations, fractional calculus and shift operator}
\author{D. Babusci$^\dag$, G. Dattoli$^\ddag$, D. Sacchetti$^\diamond$} 

\address{$^\dag$ INFN - Laboratori Nazionali di Frascati, via E. Fermi 40, I-00044 Frascati.}
\email{danilo.babusci@lnf.infn.it}

\address{$^\ddag$ ENEA - Dipartimento Tecnologie Fisiche e Nuovi Materiali, Centro Ricerche Frascati\\
                 C. P. 65, I-00044 Frascati.}
 \email{giuseppe.dattoli@enea.it}
                 
\address{$^\diamond$ Dipartimento di Statistica, Probabilit\`a e Statistica Applicata, Universit\`a 
                  "Sapienza" di Roma, P.le A. Moro, 5, 00185 Roma.}
\email{dario.sacchetti@uniroma1.it}

\begin{abstract}
We present an extension of a previously developed method employing the formalism of the fractional derivatives to 
solve new classes of integral equations. This method uses different forms of integral operators that generalizes the 
exponential shift operator.
\end{abstract}

\maketitle
\section{Introduction}\label{sec:intro}
In a previous note \cite{Babu} we have obtained a solution of the Lamb integral equation \cite{Lamb,Mart} using a formalism 
based on the fractional derivatives. Our result coincides with the solution provided, without any proof, by Bateman \cite{Mart}. 
A generalization of our method has been proposed in \cite{Fuji}, where fractional derivatives are used to solve a more 
complicated version of the Lamb equation. 

In this note we discuss a further extension of the technique put forward in \cite{Babu} and we will develop a more general 
procedure to treat integral equations whose solution requires fractional forms of differential operators other than the ordinary 
derivative. To this aim we remind that the exponential operator
\begin{equation}
\hat{E} (\lambda) \,=\, \exp \left\{\lambda\,x\,\partial_x\right\}\,,
\end{equation}
is a dilatation operator whose action on a given function $g(x)$ is (see ref. \cite{Datt1})
\begin{equation}
\hat{E} (\lambda)\,g(x) \,=\, g(\mathrm{e}^\lambda x)\;.
\end{equation}

Let us now consider the operator  $(x\,\partial_x)^{-\nu}$ with $\Re\,\nu > 0$. The use of the following property of the Laplace 
transform
\begin{equation}
a^{-\nu} \,=\, \frac{1}{\Gamma (\nu)}\,\int_0^\infty\,\mathrm{d}s \, \mathrm{e}^{- s a}\,s^{\nu - 1}\;,
\end{equation}
allows to write
\begin{equation}
(x\,\partial_x)^{-\nu} \,=\, \frac{1}{\Gamma (\nu)}\,\int_0^\infty\,\mathrm{d}s \, \hat{E}(-s)\,s^{\nu - 1}\;,
\end{equation}
and, according to the previous discussion, we find \cite{Datt2}
\begin{eqnarray}
\label{eq:genRL}
(x\,\partial_x)^{-\nu}\,g(x) \!\!&=&\!\!  \frac{1}{\Gamma (\nu)}\,\int_0^\infty\,\mathrm{d}s \, \hat{E}(-s)\,g(x)\,s^{\nu - 1} \\
   &=&\!\!  \frac{1}{\Gamma (\nu)}\, \int_0^x\,\mathrm{d}\xi\,\frac{g(\xi)}{\xi}\,\ln^{\nu - 1}\left(\frac{x}{\xi}\right)\;, \nonumber
\end{eqnarray}
which is essentially the generalization of the Riemann-Liouville integral representation for non-integer negative powers 
of the operator  $x\,\partial_x$ \cite{OldSpa}.

\section{Integral equations and fractional differential operators}\label{sec:integ}
The procedure outlined in sec. \ref{sec:intro} can be generalized to other differential forms. The first example we consider 
is the following modified form of the Lamb-Bateman equation
\begin{equation}
\label{eq:modLB}
\int_0^\infty\,\mathrm{d}y\, u(\mathrm{e}^{- y^2} x) \,=\, f(x)\;,
\end{equation}
where $u(x)$ is the function to be determined, and $f(x)$ is a continuous, free from singularities, known function.
According to the properties of the dilatation operator, we can cast eq. \eqref{eq:modLB} in the form
\begin{equation}
\int_0^\infty\,\mathrm{d}y\, \mathrm{e}^{- y^2\,x\,\partial_x}\,u(x) \,=\, f(x)\;,
\end{equation}
and, treating the derivative operator as a generic constant, the evaluation of the the gaussian integral yields
\begin{equation}
u(x) \,=\, \frac{2}{\sqrt{\pi}}\,\sqrt{x\,\partial_x}\,f(x)\,.
\end{equation}
By rewriting the operator $\sqrt{x\,\partial_x}$ as $(x\,\partial_x)\,(x\,\partial_x)^{-1/2}$, and taking into account 
eq. \eqref{eq:genRL}, we get\footnote{Let us remark that, as a consequence of the relation $[f (\hat{O}),\,\hat{O}] = 0$, 
with $f$ generic function that admits a power series expansion,  one has:  
$(x\,\partial_x)\,(x\,\partial_x)^{-1/2} = (x\,\partial_x)^{-1/2}\,(x\,\partial_x)$.}
\begin{equation}
\label{eq:funux}
u(x) \,=\, \frac{2}{\pi}\, \int_0^x\,\mathrm{d}\xi\,\frac{\partial_\xi f(\xi)}{\sqrt{\ln (x/\xi)}}
\end{equation}
whose correctness as solution of eq. \eqref{eq:modLB} has been checked numerically.

As a further example of integral equation which does not involve fractional operators but requires the use 
of the dilatation operator, we consider the following problem
\begin{equation}
\int_0^\infty\,\mathrm{d}y\, \mathrm{e}^{-y}\,u(y\,x) \,=\, f(x)\;,
\end{equation}
where $u(x)$ is the unknown function. In this case we get 
\begin{equation}
f(x) \,=\, \int_0^\infty\,\mathrm{d}y\, \mathrm{e}^{-y}\,y^{x\,\partial_x} u(x) \,=\, \Gamma (x\,\partial_x + 1)\,u(x)\;,
\end{equation}
and, thus
\begin{equation}
u(x) \,=\, \left\{\Gamma (x\,\partial_x + 1)\right\}^{-1}\,f(x)\;.
\end{equation}

In the case $f (x) = \mathrm{e}^{-x}$, we obtain the solution in terms of 0-th order Bessel functions of first kind, 
namely\footnote{Note that from $(x \partial_x) x^n = n x^n$ we obtain the following identity $g(x\,\partial_x) x^n = g(n) x^n$.}
\begin{eqnarray}
u (x) \!\!&=&\!\! \sum_{n = 0}^\infty\, \frac{(-)^n}{n!}\, \left\{\Gamma (x\,\partial_x + 1)\right\}^{-1}\,x^n \\
             &=&\!\!  \sum_{n = 0}^\infty\, \frac{(-)^n}{(n!)^2}\,x^n\,=\, J_0 (2 \sqrt{x})\;. \nonumber
\end{eqnarray}
In general, if  $f(x)$ is any function which can be expanded as
\begin{equation}
f (x) \,=\, \sum_{n = 0}^\infty\, \frac{a_n}{n!}\,x^n\;,
\end{equation}
the solution writes
\begin{equation}
u (x) \,=\, \sum_{n = 0}^\infty\, \frac{a_n}{(n!)^2}\,x^n\;.
\end{equation}
It is interesting to note that in the case in which the equation is of the type
\begin{equation}
\int_0^\infty\,\mathrm{d}y\, \mathrm{e}^{-y}\,u(y^m x) \,=\, \mathrm{e}^{-x}\;,
\end{equation}
the relevant solution can be written in terms of Bessel-like functions as follows
\begin{eqnarray}
u (x) \!\!&=&\!\! \sum_{n = 0}^\infty\, \frac{(-)^n}{n!}\, \left\{\Gamma (m\,x\,\partial_x + 1)\right\}^{-1}\,x^n \\
             &=&\!\!  \sum_{n = 0}^\infty\, \frac{(-)^n}{n!\,(m\,n)!}\,x^n\,=\, W_0 (x | m)\;. \nonumber
\end{eqnarray}
where
\begin{equation}
W_n (x | m) \,=\, \sum_{k = 0}^\infty\, \frac{(-)^k}{k! (m\,k + n)!}\,x^k
\end{equation}
is the Bessel-Wright function of order $n$ \cite{EMOT}. The same result holds for the integer $m$ replaced by any real 
$\mu$ .

We go back to fractional operational calculus by considering the following example of integral equation
\begin{equation}
\label{eq:morLB}
\int_0^\infty\,\mathrm{d}y\,u(\sqrt{x^2 + 2 y^2}) \,=\, f(x)
\end{equation}
which is apparently more complicated than the original Lamb-Bateman equation, but, as will be shown later, in spite of its different 
form, it represents the same mathematical problem. Also in this case the solution can be obtained using the previously outlined 
shift operator technique. We first recall the identity \cite{Datt2}
\begin{equation}
\label{eq:iden}
\exp\left\{\lambda\,\frac{1}{x}\,\partial_x\right\}\,g(x) \,=\, g(\sqrt{x^2 + 2 \lambda})
\end{equation}
and therefore, as a straightforward consequence of our procedure, we find
\begin{equation}
u(x) \,=\, - \frac{2}{\sqrt{\pi}}\,\sqrt{- \frac{1}{x}\,\partial_x}\,f(x)\;. 
\end{equation}
The use of the same logical steps leading to eq. \eqref{eq:funux} yields the following expression for the solution of eq. \eqref{eq:morLB}
\begin{equation}
u (x) \,=\, - \frac{\sqrt{2}}{\pi}\,\int_{x^2}^{\infty}\,\mathrm{d}\xi\,\frac{\partial_\xi f(\xi)}{\sqrt{\xi\,(\xi - x^2)}}\;.
\end{equation}

\section{Concluding remarks}\label{sec:conclu}
\begin{figure}
\begin{center}
\includegraphics[width=2.5in]{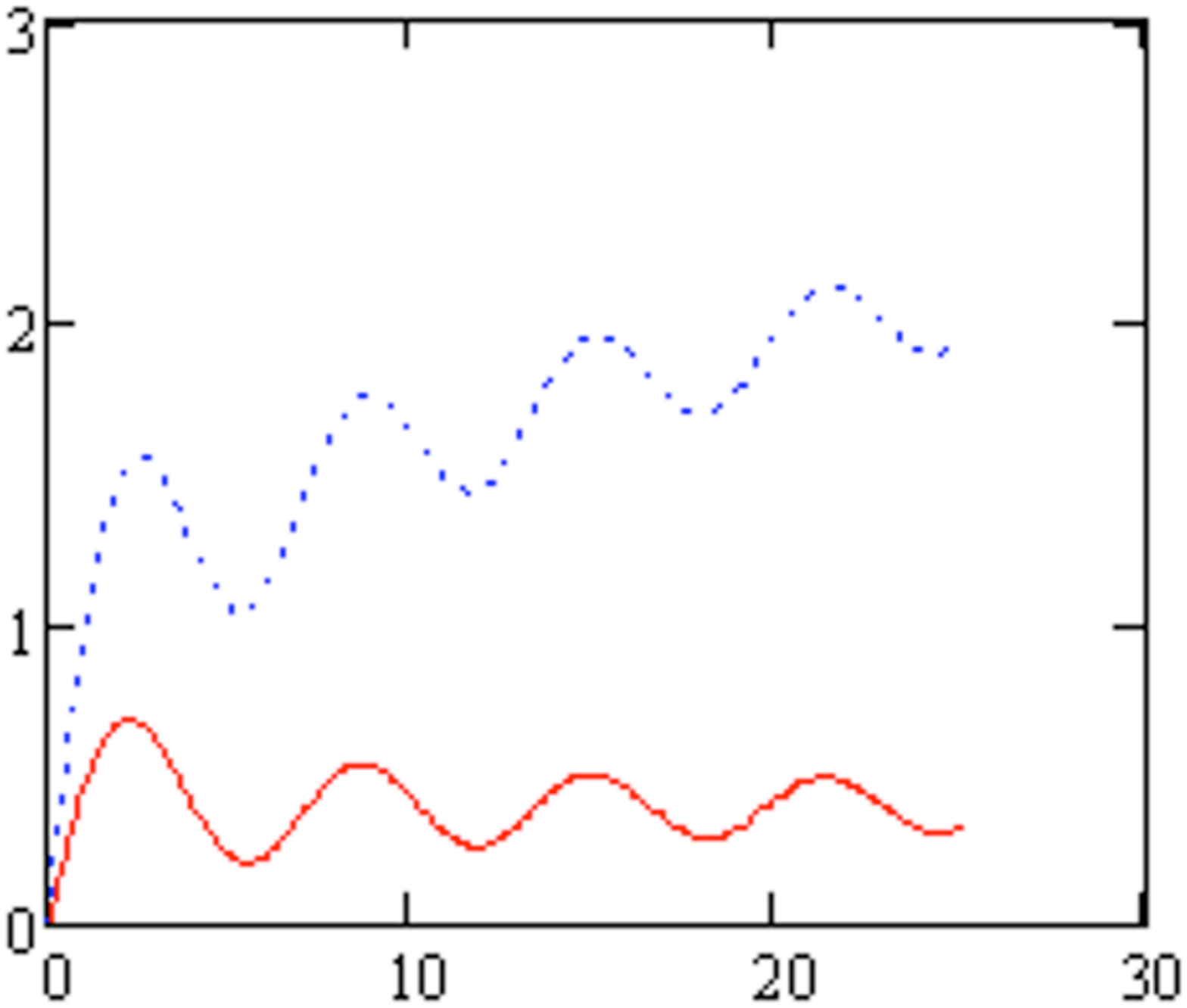}
\vspace*{-0.7in}
\caption{ The function $F (x;\nu)$ for $\nu = 1.5$ (dot) and $\nu = 4.1$ (solid).}
\label{fig:funS}
\end{center}
\end{figure}

The examples  discussed so far show that the method we have proposed is fairly powerful and is amenable of 
useful generalizations. A further example supporting the usefulness of our technique is offered by the calculation of the 
integral
\begin{equation}
\label{eq:intS}
F (x;\nu) \,=\, \int_0^\infty\,\mathrm{d}t\,\sin\left[\frac{x}{(1 + t^2)^\nu}\right]\; \qquad \qquad \qquad
\left(\Re \nu\,\geq\,\frac{1}{2}\right)\;.
\end{equation}
The shift operator method outlined in the introductory remarks allows to write this integral as follows
\begin{equation}
F (x;\nu) \,=\, \hat{O}_\nu\,\sin x \qquad \qquad \mathrm{with}\qquad \hat{O}_\nu \,=\, \int_0^\infty\,\mathrm{d}t\,
\left(\frac{1}{1 + t^2}\right)^{\nu\,x\,\partial_x}\;.
\end{equation}
The operator $\hat{O}_\nu$ can be defined explicitly by using the following result \cite{EMOT}
\begin{equation}
 \int_0^\infty\,\mathrm{d}y\,\frac{1}{(1 + y^2)^\mu} \,=\, \frac{\sqrt{\pi}}{2}\,\frac{\Gamma(\mu - 1/2)}{\Gamma(\mu)}\;,
\end{equation}
that specified to the case $\mu = \nu x \partial_x$, yields
\begin{equation}
F (x;\nu) \,=\, \frac{\sqrt{\pi}}{2}\,\frac{\Gamma(\nu\,x\,\partial_x - 1/2)}{\Gamma(\nu\,x\,\partial_x)}\,\sin x\;.
\end{equation}
By substituting to $\sin x$ its Taylor expansion, and assuming that the sum and the operator $\hat{O}_\nu$ can be 
interchanged, we get
\begin{eqnarray}
\label{eq:exprS}
F (x;\nu) \!\!&=&\!\! \frac{\sqrt{\pi}}{2}\,\Gamma(\nu\,x\,\partial_x - 1/2)\,\left\{\Gamma(\nu\,x\,\partial_x)\right\}^{-1}\,
                                \sum_{n = 0}^\infty\,\frac{(-)^n}{(2 n + 1)!}\,x^{2 n + 1} \\
                   &=&\!\! \frac{\sqrt{\pi}}{2}\, \sum_{n = 0}^\infty\,\frac{(-)^n}{(2 n + 1)!}\,
                               \frac{\Gamma\left((2 n + 1)\nu - 1/2\right)}{\Gamma\left((2 n + 1)\nu\right)}\,x^{2n + 1}\;. \nonumber
\end{eqnarray}
We have checked the correctness of this result performing a direct numerical integration of eq. \eqref{eq:intS}.  The behavior of the 
functions $F (x;\nu)$ for different values of $\nu$ is shown in Fig. \ref{fig:funS}. 

The procedure employed for the evaluation of the integral \eqref{eq:intS} is quite general and can be extended to any integral 
of the type
\begin{equation}
\label{eq:inteI}
G (x) \,=\, \int_0^\infty\,\mathrm{d}t\,f(x\,g(t))\;
\end{equation}
In fact, writing this integral as follows 
\begin{equation}
G (x) \,=\, \int_0^\infty\,\mathrm{d}t\,\left[g(t)\right]^{x\,\partial_x}\,f(x)\;,
\end{equation}
we can calculate
\begin{equation} 
O (\mu) \,=\, \int_0^\infty\,\mathrm{d}t\,\left[g(t)\right]^\mu\;,
\end{equation}
and, thus
\begin{equation}
G (x) \,=\, \hat{O} (x\,\partial_x)\,f(x)\;.
\end{equation}

The use of the shift operator method allows also the evaluation of the following integral
\begin{equation}
I (x) \,=\, \int_0^\infty\,\mathrm{d}t\,f(\sqrt{x^2 - 2 g(t)})\;,
\end{equation}
which, on account of eq. \eqref{eq:iden}, takes the following operational form 
\begin{equation}
I (x) \,=\, \hat{Q} \left(\frac{1}{x}\,\partial_x\right)\,f(x)\;.
\end{equation}
where
\begin{equation}
Q (\mu) \,=\,  \int_0^\infty\,\mathrm{d}t\,\mathrm{e}^{- \mu\,g(t)}\;.
\end{equation}

A complete theory of the generalized shift operators has been presented in ref. \cite{Datt1}, where, among the other 
things, the following identity has been proved
\begin{equation}
\label{eq:shift}
\exp \left\{\lambda\,q(x)\,\partial_x\right\}\,g(x) \,=\, g\left[F^{- 1}\,(\lambda + F(x))\right] \quad \quad 
F(x) \,=\, \int^x\,\mathrm{d}\xi\,\frac{1}{q(\xi)}\;,
\end{equation}
which is a generalization of the ordinary shift operator (the examples discussed before are just a particular case of this formula).  
According to this identity and to the methods here discussed, it is easy to show that the solution of the following integral equation
\begin{equation}
\int_0^\infty\,\mathrm{d}y\,u\left[F^{- 1}\,(-y^2 + F(x))\right] \,=\, f(x)
\end{equation}
can be written as
\begin{equation}
u (x) \,=\, \frac{2}{\sqrt{\pi}}\,\sqrt{q(x)\,\partial_x}\,f(x)\;.
\end{equation}
In a forthcoming investigation we will discuss the conditions under which the previous solution holds and some examples of applications 
of the obtained identities.

Before closing this paper we consider the following integral equation
\begin{equation}
\label{eq:inteq}
\int_0^a\,\mathrm{d}y\,u\left(\frac{x}{1 + x y}\right) \,=\, f(x)\;.
\end{equation}
By using eq. \eqref{eq:shift}, it is easy to show that\footnote{This relation can also be obtained setting $x = 1/\xi$ 
(with $x y \neq -1$). }
\begin{equation}
u\left(\frac{x}{1 + x y}\right) \,=\, \exp\left\{-y\,x^2\,\partial_x\right\}\,u(x) 
\end{equation}
and, therefore, for the solution of eq. \eqref{eq:inteq} we obtain
\begin{equation}
u (x) \,=\, \frac{x^2\,\partial_x}{1 - \exp\left\{-a\,x^2\,\partial_x\right\}}\,f(x)
\end{equation}

In this paper we have studied the properties of differential operators like $x\,\partial_x$ and their fractional generalizations. It is well 
known that, for $n$ integer different from zero, the following identity holds \cite{Horz}
\begin{equation}
\label{eq:Stir}
(x\,\partial_x)^n \,=\, \sum_{k = 0}^n\,S (n,k)\,x^k\,\partial_x^k\;,
\end{equation}
where
\begin{equation}
S (n,k) \,=\, \frac{1}{k!}\,\sum_{j = 0}^k\,(-)^{k - j}\,\binom{k}{j}\,j^n
\end{equation}
are the Stirling number of second kind. The extension of this equation to the non-integer powers of $x\,\partial_x$ does 
not exist. We make the conjecture that in this case eq. \eqref{eq:Stir} can be modified as follows
\begin{equation}
\label{eq:Stirnu}
(x\,\partial_x)^\nu \,=\, \sum_{k = 0}^\infty\,S(k,n)\,x^k\,\partial_x^k \qquad\qquad (0\,<\,\nu\,<\,1)
\end{equation}
The numerical check confirms the validity of our conjecture, but we did not succeed in getting a rigorous proof for it. 
Eq. \eqref{eq:Stirnu}, as well as other identities presented in this paper,  will be more throughly discussed in a more general 
forthcoming paper.
\newpage

\vspace{1.0cm}


\begin{thebibliography}{99}

\bibitem{Babu} 
D. Babusci, G. Dattoli, and D. Sacchetti, 
\texttt{arXiv:1006.0184v2 [math-ph]}.

\bibitem{Lamb}
H. Lamb, 
\emph{On the diffraction of a solitary wave},
Proc. London Math. Soc. {\bf 8}, 422 (1910).

\bibitem{Mart}
P. A. Martin,
\emph{Harry Bateman: from Manchester to Manuscript Project}, \\
(www.ima.org.uk/.../mt\_april10\_harry\_bateman\_from\_manchester\_to\_manuscript\_project.pdf)

\bibitem{Fuji}
K. Fujii,
\texttt{arXiv:1006.2424v1 [math-ph]}.

\bibitem{Datt1}
G. Dattoli, P. L. Ottaviani, A. Torre, and L. Vasquez, 
Riv. Nuovo Cim. {\bf 20}, 1 (1997).

\bibitem{Datt2}
G. Dattoli, P. E. Ricci, and D. Sacchetti, 
Appl. Math and Comp. {\bf 141},  215 (2003);
D. Babusci and G. Dattoli, 
\texttt{arXiv:0911.2098v1 [math-ph]}.

\bibitem{OldSpa}
K. B. Oldham and J. Spanier
\emph{The Fractional Calculus},
Dover Publications (2006).

\bibitem{EMOT}
A. Erd\'elyi, W. Magnus, F. Oberhettinger, and F. Tricomi, 
\emph{Higher Trascendental Functions}, 
Mc Graw-Hill, New York (1953).

\bibitem{Horz}
A. Horzela et al.,
\texttt{arXiv:quant-ph/0409152}.

\end{thebibliography}
\end{document}